\documentclass[english]{article}
\usepackage{times}
\usepackage[T1]{fontenc}
\usepackage[latin1]{inputenc}
\usepackage{graphicx}

\makeatletter

\usepackage{babel}
\makeatother
\begin{document}
\setlength{\unitlength}{1cm} 

\begin{picture}(5,1)(0.46,-2.5) \put(0,0){{\it Foundations of Physics Letters} {\bf 17} (2004) pp. 479--496.} \end{picture}

\bigskip{}
\noindent \textbf{\Large ON THE MEANING OF LORENTZ COVARIANCE}{\Large \par}
\vspace{1.5cm}

~~~~~\begin{minipage}[c]{0.80\columnwidth}%
\textbf{László E. Szabó}
\medskip{}

\emph{Theoretical Physics Research Group of the Hungarian Academy
of Sciences}

\emph{Department of History and Philosophy of Science}

\emph{Eötvös University, Budapest, Hungary}

\emph{E-mail: leszabo@hps.elte.hu}\end{minipage}%

\vspace{20mm}
\noindent In classical mechanics, Galilean covariance and the principle
of relativity are completely equivalent and hold for all possible
dynamical processes. In contrast, in relativistic physics the situation
is much more complex. It will be shown that Lorentz covariance and
the principle of relativity are not completely equivalent. The reason
is that the principle of relativity actually only holds for the equilibrium
quantities that characterize the equilibrium state of dissipative
systems. In the light of this fact it will be argued that Lorentz
covariance should not be regarded as a fundamental symmetry of the
laws of physics.
\medskip{}

\vspace{10mm}

\section{\noindent INTRODUCTION}

It is a widely accepted view that special relativity, beyond its claim
about space and time (cf. {[}1{]}), is a theory providing a powerful
method for the physics of objects moving at constant velocities. The
basic idea is the following: Consider a physical object at rest in
an arbitrary inertial frame $K$. Assume we know the relevant physical
equations and know the solution of the equations describing the physical
properties of the object in question when it is at rest. All these
things are expressed in the terms of the space and time coordinates
$x_{1},x_{2},x_{3},t$ and some other quantities defined in $K$ on
the basis of $x_{1},x_{2},x_{3},t$. We now inquire as to the same
physical properties of the same object when it is, as a whole, moving
at a given constant velocity relative to $K$. In other words, the
issue is how these physical properties are modified when the object
is in motion. \underbar{}The standard method for solving this problem
is based on the \emph{relativity principle/Lorentz covariance}. It
follows from the covariance of the laws of nature relative to Lorentz
transformations that the same equations hold for the primed variables
$x'_{1},x'_{2},x'_{3},t',\ldots$ defined in the co-moving inertial
frame $K'$. Moreover, since the moving object is at rest in the co-moving
reference frame $K'$, it follows from the relativity principle that
the same rest-solution holds for the primed variables. Finally, we
obtain the solution describing the system moving as a whole at constant
velocity by expressing the primed variables through the original $x_{1},x_{2},x_{3},t,\ldots$
of $K$, applying the Lorentz transformation. 

This is the way we usually solve problems such as the electromagnetic
field of a moving point charge, the Lorentz deformation of a rigid
body, the loss of phase suffered by a moving clock, the dilatation
of the mean life of a cosmic ray $\mu$-meson, etc.

In this paper I would like to show that this method, in general, is
not correct; the system described by the solution so obtained is not
necessarily identical with the original system set in collective motion.
The reason is, as will be shown, that Lorentz covariance in itself
does not guarantee that the physical laws in question satisfy the
relativity principle in general. The principle of relativity actually
only holds for the equilibrium quantities characterizing the equilibrium
state of dissipative systems.

\section{\noindent THE RELATIVITY PRINCIPLE}

\noindent Consider two inertial frames of reference $K$ and $K'$.
Assume that $K'$ is moving at constant velocity $v$ relative to
$K$ along the axis of $x$. Assume that laws of physics are know
and empirically confirmed in inertial frame $K$, including the laws
describing the behavior of physical objects in motion relative to
$K$. Denote $x(A),y(A),z(A),t(A)$ the space and time tags of an
event $A$, obtainable by means of measuring-rods and clocks at rest
relative to $K$, and denote $x'(A),y'(A),z'(A),t'(A)$ the similar
data of the same event, obtainable by means of measuring-rods and
clocks co-moving with $K'$. In the approximation of classical physics
($v\ll c$), the relationship between $x'(A),y'(A),z'(A),t'(A)$ and
$x(A),y(A),z(A),t(A)$ can be described by the Galilean transformation:
\begin{eqnarray}
t'(A) & = & t(A)\textrm{,}\label{eq:Gelso1}\\
x'(A) & = & x(A)-v\, t(A)\textrm{,}\\
y'(A) & = & y(A)\textrm{,}\\
z'(A) & = & z(A)\textrm{.}\label{eq:Gelso2}\end{eqnarray}
Due to the relativistic deformations of measuring-rods and clocks,
the exact relationship is described by the Lorentz transformation:\begin{eqnarray}
t'(A) & = & \frac{t(A)-\frac{v\, x(A)}{c^{2}}}{\sqrt{1-\frac{v^{2}}{c^{2}}}}\textrm{,}\label{eq:Lorentz1}\\
x'(A) & = & \frac{x(A)-v\, t(A)}{\sqrt{1-\frac{v^{2}}{c^{2}}}}\textrm{,}\\
y'(A) & = & y(A)\textrm{,}\\
z'(A) & = & z(A)\textrm{.}\label{eq:Lorentz2}\end{eqnarray}

Since physical quantities are defined by the same operational procedure
in all inertial frame of reference, the transformation rules of the
space and time coordinates (usually) predetermine the transformations
rules of the other physical variables. So, depending on the context,
we will mean by Galilean/Lorentz transformation not only the transformation
of the space and time tags, but also the corresponding transformation
of the other variables in question.

Following Einstein's 1905 paper, the Lorentz transformation rules
are usually derived from the relativity principle---the general validity
of which we are going to challenge in this paper. As we will see,
this derivation is not in contradiction with our final conclusions.
However, it is worth while to mention that Lorentz transformation
can also be derived independently of the principle of relativity,
directly from the facts that a clock slows down by factor $\sqrt{1-v^{2}/c^{2}}$
when it is gently accelerated from $K$ to $K'$ and a measuring-rod
suffers a contraction by factor $\sqrt{1-v^{2}/c^{2}}$ when it is
gently accelerated from $K$ to $K'$ (see {[}1{]}). 

Now, relativity principle is the following assertion: 

\medskip{}
\noindent \textbf{Relativity Principle}\emph{~~~The behavior of
the system co-moving as a whole with} $K'$\emph{, expressed in terms
of the results of measurements obtainable by means of measuring-rods,
clocks, etc., co-moving with} $K'$ \emph{is the same as the behavior
of the original system, expressed in terms of the measurements with
the equipments at rest in} $K$\emph{. }
\medskip{}

\noindent Exactly as Galilei describes it: 

\begin{quote}
... the butterflies and flies will continue their flights indifferently
toward every side, nor will it ever happen that they are concentrated
toward the stern, as if tired out from keeping up with the course
of the ship, from which they will have been separated during long
intervals by keeping themselves in the air. And if smoke is made by
burning some incense, it will be seen going up in the form of a little
cloud, remaining still and moving no more toward one side than the
other. \emph{The cause of all these correspondences of effects is
the fact that the ship's motion is common to all the things contained
in it} {[}my italics{]}, and to the air also. ({[}2{]}, p. 187)
\end{quote}
Or, in Einstein's formulation: 

\begin{quote}
If, relative to $K$, $K'$ is a uniformly moving co-ordinate system
devoid of rotation, then natural phenomena run their course with respect
to $K'$ according to exactly the same general laws as with respect
to $K$. ({[}3{]}, p. 16)
\end{quote}
In classical physics, the space and time tags obtained by means of
measuring-rods and clocks co-moving with different inertial reference
frames can be connected through the Galilean transformation. According
to special relativity, the space and time tags obtained by means of
measuring-rods and clocks co-moving with different inertial reference
frames are connected through the Lorentz transformation. Consequently,
the laws of physics must preserve their forms with respect of the
Galilean/Lorentz transformation. Thus, it must be emphasized, the
Galilean/Lorentz covariance is a \emph{consequence} of two physical
facts: 1)~the laws of physics satisfy the relativity principle and
2)~the space and time tags in different inertial frames are connected
through the Galilean/Lorentz transformation.

Let us try to unpack these verbal formulations in a more mathematical
way. Let $\mathcal{E}$ be a set of differential equations describing
the behavior of the system in question. Let us denote by $\psi$ a
typical set of (usually initial) conditions determining a unique solution
of $\mathcal{E}$. Let us denote this solution by $\left[\psi\right]$.
Denote $\mathcal{E}'$ and $\psi'$ the equations and conditions obtained
from $\mathcal{E}$ and $\psi$ by substituting every $x_{i}$ with
$x'_{i}$, and $t$ with $t'$, etc. Denote $G_{v}\left(\mathcal{E}\right),G_{v}\left(\psi\right)$
and $\Lambda_{v}\left(\mathcal{E}\right),\Lambda_{v}\left(\psi\right)$
the set of equations and conditions expressed in the primed variables
applying the Galilean and the Lorentz transformations, respectively
(including, of course, the Galilean/Lorentz transformations of all
other variables different from the space and time coordinates). Finally,
in order to give a strict mathematical formulation of relativity principle,
we have to fix two further concepts, the meaning of which are vague:
Let a solution $\left[\psi_{0}\right]$ is stipulated to describe
the behavior of the system when it is, as a whole, at rest relative
to $K$. Denote $\psi_{v}$ the set of conditions and $\left[\psi_{v}\right]$
the corresponding solution of $\mathcal{E}$ that are stipulated to
describe the similar behavior of the system as $\left[\psi_{0}\right]$
but, in addition, when the system was previously set, as a whole,
into a collective translation at velocity $v$. 

Now, what relativity principle states is \emph{equivalent} to the
following:\begin{eqnarray}
G_{v}\left(\mathcal{E}\right) & = & \mathcal{E}'\textrm{,}\label{eq:Gequivalent1}\\
G_{v}\left(\psi_{v}\right) & = & \psi'_{0}\textrm{,}\label{eq:Gequivalent2}\end{eqnarray}
in the case of classical mechanics, and \begin{eqnarray}
\Lambda_{v}\left(\mathcal{E}\right) & = & \mathcal{E}'\textrm{,}\label{eq:Lequivalent1}\\
\Lambda_{v}\left(\psi_{v}\right) & = & \psi'_{0}\textrm{,}\label{eq:Lequivalent2}\end{eqnarray}
in the case of special relativity. 

Although relativity principle implies Galilean/Lorentz covariance,
the relativity principle, as we can see, \emph{is not equivalent}
to the Galilean covariance (\ref{eq:Gequivalent1}) in itself or the
Lorentz covariance (\ref{eq:Lequivalent1}) in itself. It is equivalent
to the satisfaction of (\ref{eq:Gequivalent1}) in conjunction with
condition (\ref{eq:Gequivalent2}) in classical physics, or (\ref{eq:Lequivalent1})
in conjunction with (\ref{eq:Lequivalent2}) in relativistic physics. 

Note, that $\mathcal{E}$, $\psi_{0}$, and $\psi_{v}$ as well as
the transformations $G_{v}$ and $\Lambda_{v}$ are given by contingent
facts of nature. It is therefore a contingent fact of nature whether
a certain law of physics is Galilean or Lorentz covariant, and, \emph{independently},
whether it satisfies the principle of relativity. The relativity principle
and its consequence the principle of Lorentz covariance are certainly
normative principles in contemporary physics, providing a heuristic
tool for constructing new theories. We must emphasize however that
these normative principles, as any other fundamental law of physics,
are based on empirical facts; they are based on the observation that
the behavior of any moving physical object satisfies the principle
of relativity. In the rest of this paper I will show, however, that
the laws of relativistic physics, in general, do not satisfy this
condition. 

Before we begin analyzing our examples, it must be noted that the
major source of confusion is the vagueness of the definitions of conditions
$\psi_{0}$ and $\psi_{v}$. In principle any $\left[\psi_{0}\right]$
can be considered as a {}``solution describing the system's behavior
when it is, as a whole, at rest relative to $K$''. Given any one
fixed $\psi_{0}$, it is far from obvious, however, what is the corresponding
$\psi_{v}$. When can we say that $\left[\psi_{v}\right]$ describes
the similar behavior of the same system when it was previously set
into a collectives motion at velocity $v$? As we will see, there
is an unambiguous answer to this question in the Galileo covariant
classical physics. But $\psi_{v}$ is vaguely defined in relativity
theory. Note that Einstein himself uses this concept in a vague way.
Consider the following example:

\begin{quote}
Let there be given a stationary rigid rod; and let its length be $l$
as measured by a measuring-rod which is also stationary. We now imagine
the axis of the rod lying along the axis of $x$ of the stationary
system of co-ordinates, and that a uniform motion of parallel translation
with velocity $v$ along the axis of $x$ in the direction of increasing
$x$ is then imparted to the rod. We now inquire as to the length
of the moving rod, and imagine its length to be ascertained by the
following two operations:
\begin{itemize}
\item [(a)] The observer moves together with the given measuring-rod and
the rod to be measured, and measures the length of the rod directly
by superposing the measuring-rod, in just the same way as if all three
were at rest.
\item [(b)] ...
\end{itemize}
In accordance with the principle of relativity the length to be discovered
by the operation (a)--we will call it {}``the length of the rod in
the moving system''--must be equal to the length $l$ of the stationary
rod. {[}4{]}

\end{quote}
But, what exactly does {}``a uniform motion of parallel translation
with velocity $v$ ... imparted to the rod'' mean? The following
examples will illustrate that the vague nature of this concept complicates
matters.

In all examples we will consider a set of interacting particles. We
assume that the relevant equations describing the system are Galilean/Lorentz
covariant, that is (\ref{eq:Gequivalent1}) and (\ref{eq:Lequivalent1})
are satisfied respectively. As it follows from the covariance of the
corresponding equations, $G_{v}^{-1}\left(\psi'_{0}\right)$ and,
respectively, $\Lambda_{v}^{-1}\left(\psi'_{0}\right)$ are conditions
determining new solutions of $\mathcal{E}$. The question is whether
these new solutions are identical with the one determined by $\psi_{v}$.
If so then the relativity principle is satisfied.

\begin{figure}
\begin{center}\includegraphics[%
  scale=0.5]{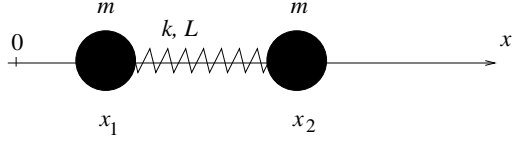}\end{center}

\caption{\emph{Two point masses are connected with a spring of equilibrium
length} $L$ \emph{and of spring constant} $k$\label{cap:Two-point-masses}}
\end{figure}

Let us start with an example illustrating how the relativity principle
works in classical mechanics. Consider a system consisting of two
point masses connected with a spring (Fig.~\ref{cap:Two-point-masses}).
 The equations of motion in $K$,\begin{eqnarray}
m\frac{d^{2}x_{1}\left(t\right)}{dt^{2}} & = & k\left(x_{2}\left(t\right)-x_{1}\left(t\right)-L\right)\textrm{,}\label{eq:mozg1}\\
m\frac{d^{2}x_{2}\left(t\right)}{dt^{2}} & = & -k\left(x_{2}\left(t\right)-x_{1}\left(t\right)-L\right)\textrm{,}\label{eq:mozg2}\end{eqnarray}
are indeed covariant with respect to the Galilean transformation,
that is, expressing (\ref{eq:mozg1})--(\ref{eq:mozg2}) in terms
of variables $x',t'$ they have exactly the same form as before:\begin{eqnarray}
m\frac{d^{2}x'_{1}\left(t'\right)}{dt'^{2}} & = & k\left(x'_{2}\left(t'\right)-x'_{1}\left(t'\right)-L\right)\textrm{,}\label{eq:mozg1'}\\
m\frac{d^{2}x'_{2}\left(t'\right)}{dt'^{2}} & = & -k\left(x'_{2}\left(t'\right)-x'_{1}\left(t'\right)-L\right)\textrm{.}\label{eq:mozg2'}\end{eqnarray}

Consider the solution of the (\ref{eq:mozg1})--(\ref{eq:mozg2})
belonging to an arbitrary initial condition $\psi{}_{0}$:\begin{equation}
\begin{array}{c}
\begin{array}{rcl}
x_{1}(t=0) & = & x_{10}\textrm{,}\\
x_{2}(t=0) & = & x_{20}\textrm{,}\\
\left.\frac{dx_{1}}{dt}\right|_{t=0} & = & v_{10}\textrm{,}\\
\left.\frac{dx_{2}}{dt}\right|_{t=0} & = & v_{20}\textrm{.}\end{array}\end{array}\label{eq:iniGal1}\end{equation}
The corresponding {}``primed'' initial condition $\psi'_{0}$ is\begin{equation}
\begin{array}{rcl}
x'_{1}(t'=0) & = & x_{10}\textrm{,}\\
x'_{2}(t'=0) & = & x_{20}\textrm{,}\\
\left.\frac{dx'_{1}}{dt'}\right|_{t'=0} & = & v_{10}\textrm{,}\\
\left.\frac{dx'_{2}}{dt'}\right|_{t'=0} & = & v_{20}\textrm{.}\end{array}\label{eq:iniGal2}\end{equation}
Applying the inverse Galilean transformation we obtain a set of conditions
$G_{v}^{-1}\left(\psi'_{0}\right)$ determining a new solution of
the original equations:\begin{equation}
\begin{array}{c}
\begin{array}{rcl}
x_{1}(t=0) & = & x_{10}\textrm{,}\\
x_{2}(t=0) & = & x_{20}\textrm{,}\\
\left.\frac{dx_{1}}{dt}\right|_{t=0} & = & v_{10}+v\textrm{,}\\
\left.\frac{dx_{2}}{dt}\right|_{t=0} & = & v_{20}+v\textrm{.}\end{array}\end{array}\label{eq:iniGal3}\end{equation}

One can recognize that this is nothing but $\psi{}_{v}$. It is the
set of the original initial conditions in superposition with a uniform
translation at velocity $v$. That is to say, the corresponding solution
describes the behavior of the same system when it was (at $t=0$)
set into a collective translation at velocity $v$, in superposition
with the original initial conditions.

In classical mechanics, as we have seen from this example, the equations
of motion not only satisfy the Galilean covariance, but also satisfy
the condition (\ref{eq:Gequivalent2}). The principle of relativity
holds for \emph{all details of the dynamics} of the system. There
is no exception to this rule. In other words, if the world were governed
by classical mechanics, relativity principle would be a universally
valid principle. With respect to later questions, it is worth noting
that the Galilean principle of relativity therefore also holds for
the equilibrium characteristics of the system, if the system has dissipations.
Imagine for example that the spring has dissipations during its distortion.
Then the system has a stable equilibrium state in which the equilibrium
distance between the particles is $L$. When we initiate the system
in collective motion corresponding to (\ref{eq:iniGal3}), the system
relaxes to another equilibrium state in which the distance between
the particles is the same $L$.

Let us turn now to the relativistic examples. It is widely held that
the new solution determined by $\Lambda_{v}^{-1}\left(\psi'_{0}\right)$,
in analogy to the solution determined by $G_{v}^{-1}\left(\psi'_{0}\right)$
in classical mechanics, describes a system identical with the original
one, but co-moving with the frame $K'$, and that the behavior of
the moving system, expressed in terms of the results of measurements
obtainable by means of measuring-rods and clocks co-moving with $K'$
is, due to Lorentz covariance, the same as the behavior of the original
system, expressed in terms of the measurements with the equipments
at rest in $K$---in accordance with the principle of relativity.
However, the situation is in fact far more complex, as I will now
show. 

Imagine a system consisting of interacting particles (for example,
relativistic particles coupled to electromagnetic field). Consider
the solution of the Lorentz covariant equations in question that belongs
to the following general initial conditions: \begin{eqnarray}
\mathbf{r}_{i}(t=0) & = & \mathbf{R}_{i}\textrm{,}\label{eq:kezdo1a}\\
\left.\frac{d\mathbf{r}_{i}(t)}{dt}\right|_{t=0} & = & \mathbf{w}_{i}.\label{eq:kezdo2a}\end{eqnarray}
(Sometimes the initial conditions for the particles unambiguously
determine the initial conditions for the whole interacting system.
Anyhow, we are omitting the initial conditions for other variables
which are not interesting now.) It follows from the Lorentz covariance
that there exists a solution of the {}``primed'' equations, which
satisfies the same conditions, \begin{eqnarray}
\mathbf{r}_{i}'(t'=0) & = & \mathbf{R}_{i}\textrm{,}\label{eq:kezdo1'a}\\
\left.\frac{d\mathbf{r}_{i}'(t')}{dt'}\right|_{t'=0} & = & \mathbf{w}_{i}\textrm{.}\label{eq:kezdo2'a}\end{eqnarray}
Eliminating the primes by means of the Lorentz transformation we obtain\begin{eqnarray}
t_{i}^{\star} & = & \frac{\frac{v}{c^{2}}R_{xi}}{\sqrt{1-\frac{v^{2}}{c^{2}}}}\textrm{,}\label{eq:kezdoidoka}\\
\mathbf{r}_{i}^{new}\left(t=t_{i}^{\star}\right) & = & \left(\begin{array}{c}
\frac{R_{xi}}{\sqrt{1-\frac{v^{2}}{c^{2}}}}\\
R_{yi}\\
R_{zi}\end{array}\right)\textrm{,}\label{eq:ujkezdeti1a}\end{eqnarray}
and\begin{eqnarray}
\left.\frac{d\mathbf{r}_{i}^{new}(t)}{dt}\right|_{t_{i}^{\star}} & = & \left(\begin{array}{c}
\frac{w_{xi}+v}{1+\frac{w_{xi}v}{c^{2}}}\\
w_{yi}\\
w_{zi}\end{array}\right)\textrm{.}\label{eq:ujkezdeti2a}\end{eqnarray}
It is difficult to tell what the solution deriving from such a nondescript
{}``initial'' condition is like, but it is not likely to describe
the original system in collective motion at velocity $v$. The reason
for this is not difficult to understand. Let me explain it by means
of a well known old example {[}5--9{]}: Consider the above system
consisting of two particles connected with a spring (two rockets connected
with a thread in the original example). Let us first ignore the spring.
Assume that the two particles are at rest relative to $K$, one at
the origin, the other at the point $d$, where $d=L$, the equilibrium
length of the spring when it is at rest. It follows from (\ref{eq:kezdoidoka})--(\ref{eq:ujkezdeti2a})
that the Lorentz boosted system corresponds to two particles moving
at constant velocity $v$, such that their motions satisfy the following
conditions:\begin{eqnarray}
t_{1}^{\star} & = & 0\textrm{,}\nonumber \\
t_{2}^{\star} & = & \frac{\frac{v}{c^{2}}d}{\sqrt{1-\frac{v^{2}}{c^{2}}}}\textrm{,}\nonumber \\
r_{1}^{new}\left(0\right) & = & 0\textrm{,}\nonumber \\
r_{2}^{new}\left(\frac{\frac{v}{c^{2}}d}{\sqrt{1-\frac{v^{2}}{c^{2}}}}\right) & = & \frac{d}{\sqrt{1-\frac{v^{2}}{c^{2}}}}\textrm{.}\label{eq:osszetartozok}\end{eqnarray}
However, the corresponding new solution of the equations of motion
does not {}``know'' about how the system was set into motion and/or
how the state of the system corresponding to the above conditions
comes about. Consider the following possible scenarios:

\paragraph*{Example~1}

The two particles are at rest; the distance between them %
\begin{figure}[h]
\begin{center}\includegraphics[%
  scale=0.5]{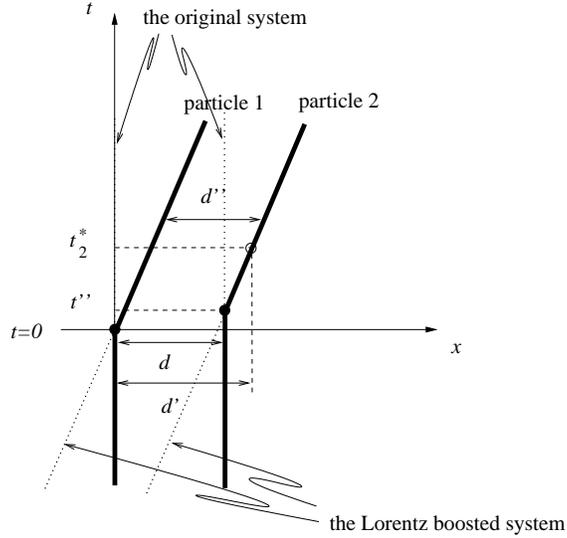}\end{center}

\caption{\emph{Both particles are at rest. Then particle 1 starts its motion
at} $t=0$. \emph{The motion of particle 2 is such that it goes through
the point} $(t_{2}^{\star},d')$\emph{, where} $d'=d/\sqrt{1-v^{2}/c^{2}}$\emph{,
consequently it started from the point of coordinate} $d$ \emph{at}
$t''=d\left(v/\left(c^{2}\sqrt{1-v^{2}/c^{2}}\right)-\left(1-\sqrt{1-v^{2}/c^{2}}\right)/\left(v\sqrt{1-v^{2}/c^{2}}\right)\right)$\emph{.
The distance between the particles at} $t''$ \emph{is} $d''=d\sqrt{1-v^{2}/c^{2}}$\emph{,
in accordance with the Lorentz contraction.}\label{cap:Both-particles-are}}
\end{figure}
 is $d$ (see Fig.~\ref{cap:Both-particles-are}). Then, particle
1 starts its motion at constant velocity $v$ at $t=0$ from the point
of coordinate $0$ (the last two dimensions are omitted); particle
2 start its motion at velocity $v$ from the point of coordinate $d$
with a delay at time $t''$. Meanwhile particle~1 moves closer to
particle~2 and the distance between them is $d''=d\sqrt{1-v^{2}/c^{2}}$,
in accordance with the Lorentz contraction. Now, one can say that
the two particles are in collective motion at velocity $v$ relative
to the original system $K$---or, equivalently, they are collectively
at rest relative to $K'$---for times $t>t_{2}^{\star}=vd/\left(c^{2}\sqrt{1-v^{2}/c^{2}}\right)$.
In this particular case they have actually been moving in this way
since $t''$. Before that time, however, the particles moved relative
to each other, in other words, the system underwent deformation.

\paragraph*{Example~2 }

\begin{figure}
\begin{center}\includegraphics[%
  scale=0.5]{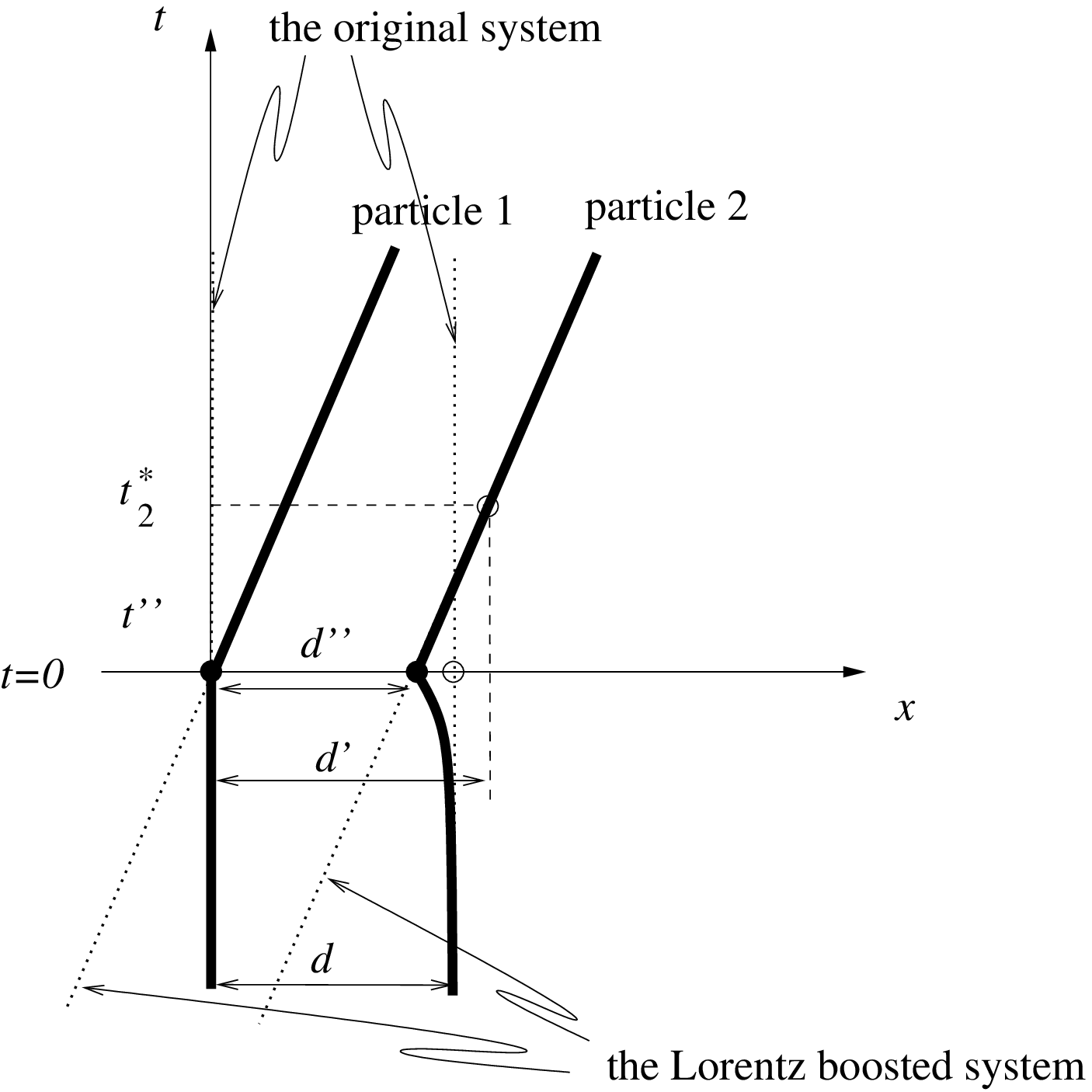}\end{center}

\caption{\emph{Both particles start at} $t=0$. \emph{Particle~2 is previously
moved to the point of coordinate} $d''=d\sqrt{1-v^{2}/c^{2}}$.\label{cap:Both-particles-start}}
\end{figure}

Both particles started at $t=0$, but particle~2 was previously moved
to the point of coordinate $d\sqrt{1-v^{2}/c^{2}}$ and starts from
there. (Fig.~\ref{cap:Both-particles-start}) %

\paragraph*{Example~3 }

Both particles started at $t=0$ %
\begin{figure}
\begin{center}\includegraphics[%
  scale=0.5]{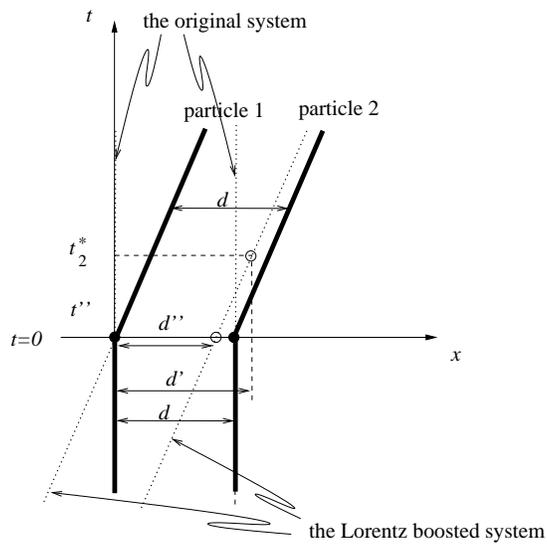}\end{center}

\caption{\emph{Both particles start at} $t=0$ \emph{from the original places.
The distance between the particles does not change.}\label{cap:Both-particles-start'}}
\end{figure}
 from their original places. The distance between them remains $d$
(Fig.~\ref{cap:Both-particles-start'}). They are in collective motion
at velocity $v$, although this motion is not described by the Lorentz
boost.

\paragraph*{Example~4 }

\begin{figure}
\begin{center}\includegraphics[%
  scale=0.5]{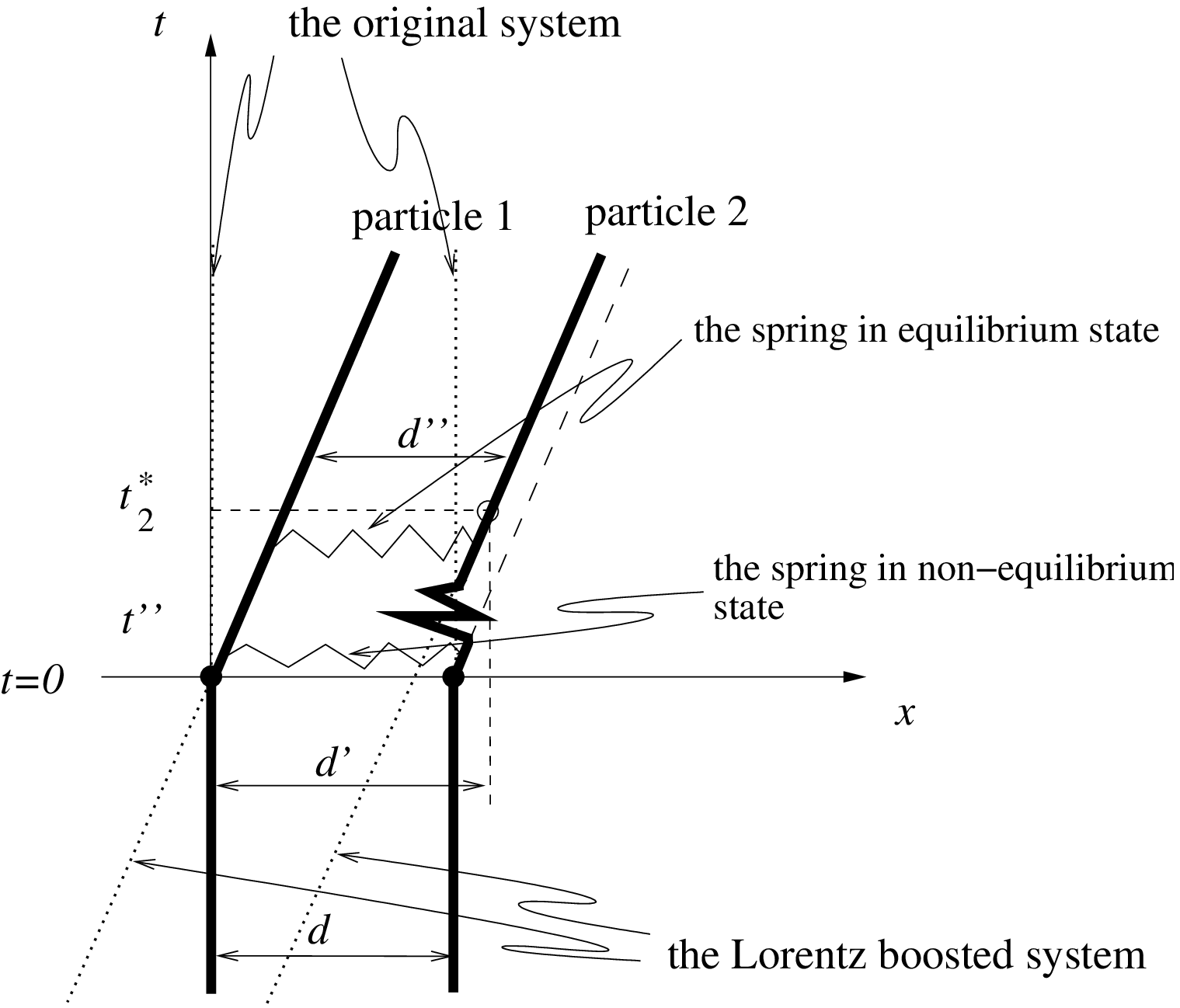}\end{center}

\caption{\emph{The particles are connected with a spring (and, say, the mass
of particle ~1 is much larger)}\label{cap:if_string}}
\end{figure}

If, however, they are connected with the spring (Fig.~\ref{cap:if_string}),
then the spring (when moving at velocity $v$) first finds itself
in a non-equilibrium state of length $d$, then it relaxes to its
equilibrium state (when moving at velocity $v$) and---assuming that
the equilibrium properties of the spring satisfy the relativity principle,
which we will argue for later on---its length (the distance of the
particles) would relax to $d\sqrt{1-v^{2}/c^{2}}$, according to the
Lorentz boost.%

We have seen from these examples that the relationship between the
Lorentz boost---the motion determined by the conditions $\Lambda_{v}^{-1}\left(\psi'_{0}\right)$---and
the systems being in collective motion---determined by $\psi{}_{v}$---is
not so trivial. In Examples 1 and 2---although, at least for large
$t$, the system is identical with the one obtained through the Lorentz
boost---it would be entirely counter intuitive to say that we simply
set the system in collective motion at velocity $v$, because we first
distorted it: in Example~1 the particles were set into motion at
different moments of time; in Example~2, before we set them in motion,
one of the particles was relocated relative to the other. In contrast,
in Examples 3 and 4 we are entitled to say that the system was set
into collective motion at velocity $v$. But, in Example~3 the system
in collective motion is different from the Lorentz boosted system
(for all $t$), while in Example~4 the moving system is indeed identical
with the Lorentz boosted one, at least for large $t$, after the relaxation
process.

Thus, as Bell rightly pointed out:

\begin{quote}
Lorentz invariance alone shows that for any state of a system at rest
there is a corresponding `primed' state of that system in motion.
But it does not tell us that if the system is set anyhow in motion,
it will actually go into the 'primed' of the original state, rather
than into the `prime' of some \emph{other} state of the original system.
({[}9{]}, p. 75)
\end{quote}
However, neither Bell's paper nor the preceding discussion of the
{}``two rockets problem'' provide a deeper explanation of this fact.
For instance, after the above passage Bell continues:

\begin{quote}
In fact, it will generally do the latter. A system set brutally in
motion may be bruised, or broken, or heated or burned. For the simple
classical atom similar things could have happened if the nucleus,
instead of being moved smoothly, had been \emph{jerked}. The electron
could be left behind completely. Moreover, a given acceleration is
or is not sufficiently gentle depending on the orbit in question.
An electron in a small, high frequency, tightly bound orbit, can follow
closely a nucleus that an electron in a more remote orbit -- or in
another atom -- would not follow at all. Thus we can only assume the
Fitzgerald contraction, etc., for a coherent dynamical system whose
configuration is determined essentially by internal forces and only
little perturbed by gentle external forces accelerating the system
as a whole. ({[}9{]}, p. 75)
\end{quote}
Of course, acceleration must be gradual so that the objects in question
are not damaged. (In our examples we omitted the acceleration period---symbolized
by a black point on the figures---for the sake of simplicity.) As
the above examples show, this condition in itself does not, however,
guarantee that the Lorentz boosted solution describes the original
system gently accelerated from $K$ to $K'$. Before I proceed to
formulate my thesis about this question, let me give one more example.

\paragraph*{Example~5 }

Consider a rod at rest in $K$. The length of the rod is $l$. At
a given moment of time $t_{0}$ we take a record about the positions
and velocities of all particles of the rod:\begin{eqnarray}
r_{i}(t=t_{0}) & = & R_{i}\textrm{,}\label{eq:kezdo1x}\\
\left.\frac{dr_{i}(t)}{dt}\right|_{t=t_{0}} & = & w_{i}\textrm{.}\label{eq:kezdo2x}\end{eqnarray}
Then, forget this system, and imagine another one which is initiated
at moment $t=t_{0}$ with the initial condition (\ref{eq:kezdo1x})--(\ref{eq:kezdo2x}).
No doubt, the new system will be identical with a rod of length $l$,
that continues to be at rest in $K$. 

Now, imagine that the new system is initiated at $t=t_{0}$ with the
initial condition\begin{eqnarray}
r_{i}(t=t_{0}) & = & R_{i}\textrm{,}\label{eq:kezdo1y}\\
\left.\frac{dr_{i}(t)}{dt}\right|_{t=t_{0}} & = & w_{i}+v\textrm{,}\label{eq:kezdo2y}\end{eqnarray}
instead of (\ref{eq:kezdo1x})--(\ref{eq:kezdo2x}). No doubt, in
a very short interval of time $\left(t_{0},t_{0}+\Delta t\right)$
this system is a rod of length $l$, moving at velocity $v$; the
motion of each particle is a superposition of its original motion,
according to (\ref{eq:kezdo1x})--(\ref{eq:kezdo2x}), and the collective
translation at velocity $v$. In other words, it is a rod co-moving
with the reference frame $K'$. Still, its length is $l$, contrary
to the principle of relativity, according to which the rod should
be of length $l\sqrt{1-v^{2}/c^{2}}$---as a consequence of $l'=l$.
The resolution of this {}``contradiction'' is that the system initiated
in state (\ref{eq:kezdo1y})--(\ref{eq:kezdo2y}) at time $t_{0}$
finds itself in a non-equilibrium state and then, due to certain dissipations,
it \emph{relaxes} to the \emph{new} equilibrium state. What such a
new equilibrium state is like, depends on the details of the dissipation/relaxation
process. It is, in fact, a \emph{thermodynamical} question. The concept
of {}``gentle acceleration'' not only means that the system does
not go irreversibly far apart from its equilibrium state, but, more
essentially, it incorporates the assumption that there is such a dissipation/relaxation
phenomenon. 

Without entering into the quantum mechanics of solid state systems,
a good way to picture it is imagine that the system is radiating during
the relaxation period. This process can be followed in details by
looking at one single point charge accelerated from $K$ to $K'$
(see {[}10{]}, pp. 208-210). Suppose the particle is at rest for $t<0$,
the acceleration starts at $t=0$ and the particle moves with constant
velocity $v$ for $t\geq t_{0}$. %
\begin{figure}
\begin{center}\includegraphics[%
  scale=0.5]{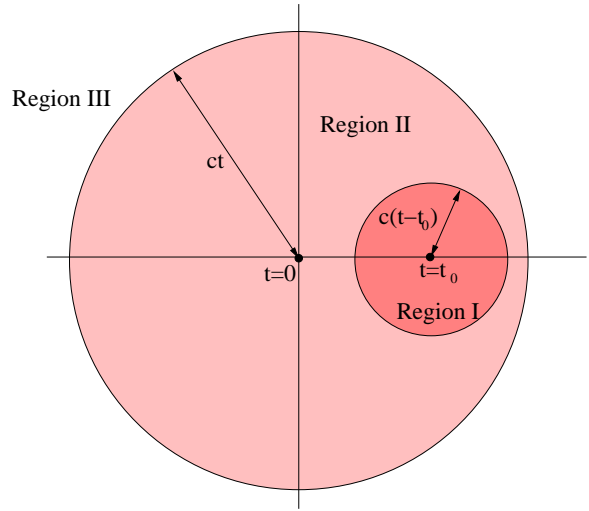}\end{center}

\caption{\emph{Scheme of regions I, II and III} \label{cap:Scheme-of-regions}}
\end{figure}
 Using the retarded potentials, we can calculate the field of the
moving particle at some time $t>t_{0}$. We find three zones in the
field (see Fig.~\ref{cap:Scheme-of-regions}). In Region I, surrounding
the particle, we find the {}``Lorentz-transformed Coulomb field''
of the point charge moving at constant velocity. This is the solution
we usually find in textbooks. In Region II, surrounding Region I,
we find a radiation field traveling outwards which was emitted by
the particle in the period $0<t<t_{0}$ of acceleration. Finally,
outside Region II, the field is produced by the particle at times
$t<0$. The field in Region III is therefore the Coulomb field of
the charge at rest. Thus, the principle of relativity \emph{never}
holds exactly. Although, the region where {}``the principle holds''
(Region I) is blowing up at the speed of light. In this way the whole
configuration relaxes to a solution which is identical with the one
derived from the principle of relativity.

Thus, we must draw the conclusion that, in spite of the Lorentz covariance
of the equations, whether or not the solution determined by the condition
$\Lambda_{v}^{-1}\left(\psi'_{0}\right)$ is identical with the solution
belonging to the condition $\psi_{v}$, in other words, whether or
not the relativity principle holds, depends on the details of the
dissipation/relaxation process in question, \emph{given that 1)~there
is dissipation in the system at all and, 2)~the physical quantities
in question, to which the relativity principle applies, are equilibrium
quantities characterizing the equilibrium properties of the system.}
For instance, in Example~5, the relativity principle does not hold
for all dynamical details of all particles of the rod. The reason
is that many of these details are sensitive to the initial conditions.
The principle holds only for some macroscopic equilibrium properties
of the system, like the length of the rod. It is a typical feature
of a dissipative system that it unlearns the initial conditions; some
of the properties of the system in equilibrium state, after the relaxation,
are independent from the initial conditions. The limiting ($t\rightarrow\infty$)
electromagnetic field of the moving charge and the equilibrium length
of a solid rod are good examples. These equilibrium properties are
completely determined by the equations themselves \emph{independently
of the initial conditions}. If so, the Lorentz covariance of the equations
in itself guarantees the satisfaction of the principle of relativity
\emph{with respect to these properties}: Let $X$ be the value of
such a physical quantity---characterizing the equilibrium state of
the system in question, fully determined by the equations independently
of the initial conditions---ascertained by the measuring devices at
rest in $K$. Let $X'$ be the value of the same quantity of the same
system when it is in equilibrium and at rest relative to the moving
reference frame $K'$, ascertained by the measuring devices co-moving
with $K'$. If the equations are Lorentz covariant, then $X=X'$.
We must recognize that whenever in relativistic physics we derive
correct results by applying the principle of relativity, we apply
it for such particular equilibrium quantities. \emph{But the relativity
principle, in general, does not hold for the whole dynamics of the
systems in relativity theory}, in contrast to classical mechanics\emph{.}

When claiming that relativity principle, in general, does not hold
for the whole dynamics of the system, a lot depends on what we mean
by the system set into uniform motion. One has to admit that this
concept is still vague. As we pointed out, it was not clearly defined
in Einstein's formulation of the principle either. By leaving this
concept vague, Einstein tacitly assumes that these details are irrelevant.
However, they can be irrelevant only if the system has dissipations
and the principle is meant to be valid only for some equilibrium properties
with respect to which the system unlearns the initial conditions.
So the best thing we can do is to keep the classical definition of
$\psi_{v}$: Consider a system of particles the motion of which satisfies
the following (initial) conditions: \begin{equation}
\begin{array}{c}
\begin{array}{rcl}
\mathbf{r}_{i}(t=t_{0}) & = & \mathbf{R}_{i0}\textrm{,}\\
\left.\frac{d\mathbf{r}_{1}}{dt}\right|_{t=t_{0}} & = & \mathbf{V}_{i0}\textrm{.}\end{array}\end{array}\label{eq:ini_classic1}\end{equation}
The system is set in collective motion at velocity $\mathbf{v}$ at
the moment of time $t_{0}$ if its motion satisfies \begin{equation}
\begin{array}{c}
\begin{array}{rcl}
\mathbf{r}_{i}(t=t_{0}) & = & \mathbf{R}_{i0}\textrm{,}\\
\left.\frac{d\mathbf{r}_{1}}{dt}\right|_{t=t_{0}} & = & \mathbf{V}_{i0}+\mathbf{v}\textrm{.}\end{array}\end{array}\label{eq:ini_classic2}\end{equation}
I have two arguments for such a choice: The usual Einsteinian derivation
of Lorentz transformation, simultaneity in $K'$, etc., starts with
the declaration of the relativity principle. Therefore, all these
things must be logically preceded by the concept of a physical object
in a uniform motion relative to $K$. The second support comes from
what Bell calls {}``Lorentzian pedagogy''.

\begin{quote}
Its special merit is to drive home the lesson that the laws of physics
in any one reference frame account for all physical phenomena, including
the observations of moving observers. And it is often simpler to work
in a single frame, rather than to hurry after each moving objects
in turn. ({[}9{]}, p. 77.)
\end{quote}

\section{\noindent CONCLUSIONS}

\noindent We have seen that in classical mechanics the principle of
relativity is, indeed, a universal principle. It holds, without any
restriction, for \emph{all} dynamical details of \emph{all} possible
systems described by classical mechanics. In contrast, in relativistic
physics this is not the case:

\begin{enumerate}
\item The principle of relativity is not a universal principle. It does
not hold for the whole range of validity of the Lorentz covariant
laws of relativistic physics, but only for the equilibrium quantities
characterizing the equilibrium state of dissipative systems. Since
dissipation, relaxation and equilibrium are thermodynamical conceptions
\emph{par excellence}, the special relativistic principle of relativity
is actually a thermodynamical principle, rather than a general principle
satisfied by all dynamical laws of physics describing all physical
processes in details. One has to recognize that the special relativistic
principle of relativity is experimentally confirmed only in such restricted
sense.
\item The satisfaction of the principle of relativity in such restricted
sense is indeed guaranteed by the Lorentz covariance of those physical
equations that determine, independently of the initial conditions,
the equilibrium quantities for which the principle of relativity holds.
In general, however, Lorentz covariance of the laws of physics does
not guarantee the satisfaction of the relativity principle.
\item It is an experimentally confirmed fact of nature that some laws of
physics are \emph{ab ovo} Lorentz covariant. However, since relativity
principle is not a universal principle, it does not entitle us to
infer that Lorentz covariance is a fundamental symmetry of physics. 
\item The fact that the space and time tags obtained by means of measuring-rods
and clocks co-moving with different inertial reference frames can
be connected through the Lorentz transformation is compatible with
our general observation that the principle of relativity only holds
for such equilibrium quantities as the length of a solid rod or the
characteristic periods of a clock-like system.
\end{enumerate}
The fact that relativity principle is not a universal principle throws
new light upon the discussion of how far the Einsteinian special relativity
can be regarded as a principle theory relative to the other (constructive)
approaches (cf. {[}11{]}, p. 57; {[}12--14{]}). It can also be interesting
from the point of view of other reflections on possible violations
of Lorentz covariance (see, for example, {[}15{]}).

It must be emphasized that the physical explanation of this more complex
picture is rooted in the physical deformations of moving measuring-rods
and moving clocks by which the space and time tags are defined in
moving reference frames. In Einstein's words:

\begin{quote}
A Priori it is quite clear that we must be able to learn something
about the physical behavior of measuring-rods and clocks from the
equations of transformation, for the magnitudes $z,y,x,t$ are nothing
more nor less than the results of measurements obtainable by means
of measuring-rods and clocks. ({[}3{]}, p. 35) 
\end{quote}
Since therefore Lorentz transformation itself is not merely a mathematical
concept without contingent physical content, we must not forget the
real physical content of Lorentz covariance and relativity principle.

\paragraph*{Acknowledgements}

The research was partly supported by the OTKA Foundation, No. T 037575
and No. T 032771. I am grateful to the Netherlands Institute for Advanced
Study (NIAS) for providing me with the opportunity, as a Fellow-in-Residence,
to complete this paper.

\section*{\noindent REFERENCES}

\begin{enumerate}
\item L. E. Szabó, {}``Does special relativity theory tell us anything
new about space and time?,'' arXiv:physics/0308035 (2003).
\item G. Galilei, \emph{Dialogue concerning the two chief world systems,
Ptolemaic \& Copernican}, (University of California Press, Berkeley,
1953).
\item A. Einstein, \emph{Relativity: The Special and General Theory} (H.
Holt, New York, 1920).
\item A. Einstein, {}``On the Electrodynamics of Moving Bodies,'' in A.
Einstein \emph{et al}, \emph{Principle of Relativity} (Dover Pubns,
London, 1924).
\item E. Dewan and M. Beran, {}``Note on Stress Effects due to Relativistic
Contraction,'' \emph{American Journal of Physics} \textbf{27}, 517
(1959).
\item A. A. Evett and R. K. Wangsness, {}``Note on the Separation of Relativistic
Moving Rockets,'' \emph{American Journal of Physics} \textbf{28},
566 (1960).
\item E. Dewan, {}``Stress Effects due to Lorentz Contraction,'' \emph{American
Journal of Physics} \textbf{31}, 383 (1963).
\item A. A. Evett, {}``A Relativistic Rocket Discussion Problem,'' \emph{American
Journal of Physics} \textbf{40}, 1170 (1972).
\item J. S. Bell, {}``How to teach special relativity,'' in \emph{Speakable
and unspeakable in quantum mechanics} (Cambridge University Press,
Cambridge, 1987).
\item L. Jánossy, \emph{Theory of relativity based on physical reality,}
(Akadémiai Kiadó, Budapest, 1971).
\item A. Einstein, {}``Autobiographical Notes,'' in \emph{Albert Einstein:
Philosopher-Scientist}, Vol. 1., P. A. Schilpp, ed. (Open Court, Illionis,
1969).
\item J. S. Bell, {}``George Francis FitzGerald,'' \emph{Physics World}
\textbf{5}, 31 (1992).
\item H. R. Brown and O. Pooley, {}``The origin of space-time metric: Bell's
'Lorentzian pedagogy' and its significance in general relativity,''
in \emph{Physics meets philosophy at the Planck scale. Contemporary
theories in quantum gravity}, C. Calleander and N. Huggett, eds. (Cambridge
University Press, Cambridge, 2001).
\item H. R. Brown, {}``The origins of length contraction: I. The FitzGerald-Lorentz
deformation hypothesis,'' \emph{American Journal of Physics} \textbf{69},
1044 (2001); {}``Michelson, FitzGerald and Lorentz: the origins of
relativity revisited,'' http://philsci-archive.pitt.edu/archive/00000987
(2003).
\item V. A. Kosteleck\'y and S. Samuel, {}``Spontaneous breaking of Lorentz
symmetry in string theory,'' \emph{Physical Review} \textbf{D39},
683 (1989).\end{enumerate}

\end{document}